# Bio-inspired hierarchical dissipative metamaterials

Marco Miniaci[1], Anastasiia Krushynska[2], Federico Bosia[2], Nicola M. Pugno[3,4,5, *]

[1]University of Le Havre, Laboratoire Ondes et Milieux Complexes, UMR CNRS 6294,

76600 Le Havre, France

[2]Department of Physics and Nanostructured Surfaces and Interfaces Centre, University of Turin,

Via Pietro Giuria 1, 10125, Torino, Italy

[3]Laboratory of Bio-Inspired & Graphene Nanomechanics, Department of Civil, Environmental and

Mechanical Engineering, University of Trento, Via Mesiano, 77, 38123, Trento, Italy

[4]School of Engineering & Materials Science, Queen Mary University of London, Mile End Road,

London, E1 4NS, UK

[5]Italian Space Agency, Via del Politecnico snc, 00133 Rome, Italy

* nicola.pugno@unitn.it

**Abstract**

Hierarchical structures with constituents over multiple length scales are found in various natural materials like bones, shells, spider silk and others, all of which display enhanced quasi-static mechanical properties, such as high specific strength, stiffness and toughness. At the same time, the role of hierarchy on the dynamic behaviour of metamaterials remains largely unexplored. This study assesses the effect of bio-inspired hierarchical organization as well as of viscoelasticity on the wave attenuation properties of continuous mechanical metamaterials. We consider single-phase metamaterials formed by self-similar unit cells with different hierarchical levels and types of hierarchy. Results highlight a number of advantages through the introduction of structural hierarchy. Band gaps relative to the corresponding non-hierarchical structures are mostly preserved, while additional "hierarchically-induced" band gaps appear. Additionally, the hierarchical configuration allows the tuning of the band gap frequencies of regular metamaterial to lower frequencies, with a simultaneous significant reduction of the global structural weight. We show that even small viscoelastic effects, not treated in the current literature, are essential in determining this behaviour. The approach we propose allows the addition of hierarchical elements to existing metamaterial configurations, with the corresponding improvement of the wave damping properties, thus providing indications for the design of structures for practical applications.

1. Introduction

Biological structural materials are renowned for their exceptional mechanical characteristics, often surpassing synthetic high-performance materials [1]. Spider silk, bone, enamel, limpet teeth are examples of materials that combine high specific strength and stiffness with outstanding toughness and flaw resistance [2, 3, 4]. Many studies have shown hierarchical structure to be responsible for these properties, e.g. providing many energy dissipation and crack deflection mechanisms over various size scales to contribute to high toughness [5]. However, studies in biomechanics linking material structure to function have mainly been limited to the quasi-static regime. The dynamic properties of these materials have been somewhat less studied, although notable examples of impact tolerance (e.g. the Bombardier Beetle's explosion chamber [6]) or vibration damping (e.g. the woodpecker skull [7]) have been observed and investigated.

A systematic study of the dynamic properties of composite structures such as those found in biomaterials requires analysis tools and methods usually adopted in the field of elastic metamaterials. The latter are periodically arranged structures with advanced dynamic characteristics, e.g. Band Gaps (BGs) within inhibited wave propagation frequencies, which enable wave focusing or collimation, negative refraction, and defect states, when the presence of a defect results in energy localization at BG frequencies. These properties make metamaterials attractive for various existing and emerging applications including seismic wave shielding [8] [9], environmental noise abatement [10, 11], vibration isolation [12], cloaking [13], imaging [14], thermal management [15], etc. In the past two decades, wave propagation mechanisms in phononic crystals based on Bragg scattering [16] and/or locally resonant metamaterials [17] have been widely studied at various spatial and frequency scales. Conventional configurations are mostly based on elementary unit cells comprising single inclusions or cavities, which results in limited BGs frequency ranges, depending on the chosen BG generation mechanisms [18, 19, 20, 21, 22]. Optimized design of the architecture of a unit cell allows to maximize the size of a single BG or even several BGs, e.g. by using topology optimization or multi-objective optimization methods [23, 24, 25]. Other metamaterial geometries comprising soft materials, which exploit thermally coupled dissipation mechanisms [26], are limited in wave attenuation performance at multiple frequency ranges, since the latter largely depends on the material thickness, preventing the design of stiff structures with strong wave shielding characteristics. Therefore, light-weight, practically oriented metamaterial structures with multi-scale wave attenuation abilities remain an open field of research.

One way to potentially overcome this limitation is to exploit fractal-like and bio-inspired structural hierarchy. Structural hierarchy is here understood in the sense that a representative metamaterial unit cell comprises multiple arrangements of inhomogeneities at various size scales. If the same arrangement occurs at every scale, the pattern is called self-similar. If a unit cell as a whole is geometrically similar to its parts at any hierarchical level, the structure is fractal-like [27]. Several recent studies have reported new classes of hierarchical and fractal-like metamaterials with superior wave attenuation in multiband frequency ranges due to their structural organization [28, 29, 30, 31, 27, 32]. These show that hierarchically organized lattice-type metamaterials exhibit multiple BGs with advanced load carrying capabilities [30], improved thermal resistance [28], and that hierarchy allows the introduction of a new type of BG generation mechanism [27]. On the other hand, continuous hierarchical metamaterials also demonstrate broadband vibration mitigation capabilities combined with improved mechanical performance [33]. These advanced properties have been attributed solely to structural hierarchy [30]. In all these cases, hierarchical designs are borrowed from those found in biological materials, e.g. the well-known "brick and mortar" microstructure [33], nacre and bone microstructures [31, 34], the biocalcite architecture [29], etc.

Some advantages of incorporating structural hierarchy in metamaterial structures have thus been highlighted for lattice and heterogeneous configurations but hierarchical single-phase continuous metamaterials remain to be investigated. A great variety of further hierarchical patterns can be found in natural structures, such as in diatoms, butterfly wings, wood, leafs, macaw feathers, grass stems, kelp, corals, cotton, human bone, cuttlefish bone, and sponges [35] [36]. All these patterns are characterized by a porous structure and can thus be exploited for bio-inspiration, providing an alternative approach to that adopted so far, which has implied the conservation of the structural mass when introducing additional hierarchical levels. Porous hierarchical structures, on the other hand, allow to fully exploit the advantages of hierarchical organization for the design of light-weight and robust continuous metamaterials.

In this work, we therefore propose to introduce hierarchy in *continuous* single-phase structures with cavities by *reducing* the unit cell weight while introducing the next hierarchical level. This approach significantly simplifies the manufacturing process and results in more *lightweight* structures compared to non-hierarchical ones, with advantages for practical applications. We study two types of hierarchical organization: a so-called "nested" hierarchy, when hierarchical objects of a subsequent level are located inside the objects of the previous level, and an "external" hierarchy, when hierarchical objects are located externally to those of the previous level. As in previous works [33, 30, 29, 34, 28, 31], the hierarchical metamaterial design here follows a bioinspired approach, meaning that we specifically draw inspiration from the structures found in porous hierarchical materials and use their organizational principles to create novel metamaterial models, although we do not aim to emulate or mimic the mechanical or dynamic properties of any specific biological system. Additionally, we provide a detailed analysis of the possibilities provided by hierarchical organization of continuous metamaterials as a function of the spatial ratio of hierarchical levels involved, and explicitly evaluate the critical effect of material viscosity.

## 2. Models and methods

Biological hierarchically-structured materials occur in a great variety of microstructural patterns [35]. In many cases, the structures are heterogeneous, typically with reinforcing (stiffer) elements such as fibres, platelets or crystals embedded in a softer, often amorphous matrix [1]. In other cases (such as in bone) voids, cavities or canals are also included, contributing to reduced mass density, but also potentially to effects on elastic wave propagation or damping [7]. Hierarchical levels (HL) often span various length scales, from nanometres to centimetres [1]. In other cases, hierarchical structures occur on similar size scales, as in the case of wood, leafs, grass stems, corals, or sponges [35] [36]. Despite the great variety of hierarchical structures occurring in nature, distinctive features can be observed and extracted, in order to drive/inspire the design of mechanical metamaterials with enhanced dynamic properties. We choose as representative structures those found in diatom cell walls (frustules) (Fig.1). Two typical hierarchical configurations are a nested arrangement of smaller scale objects inside a larger scale frame (Figs. 1a) or, in contrast, the distribution of smaller cavities around larger ones (Fig. 1b). Additionally, many biological microstructures contain high-aspect ratio (slender) constitutive elements (such as in Fig. 1a), which can play a key role in altering their vibrational characteristics.

Based on these patterns, we study metamaterial configurations with (i) thin connecting elements and a nested hierarchical organization, defined here as a "hub-spoke" geometry (Fig. 1a), and (ii) cross-like cavities with external hierarchical organization, referred to as "cross-like porous" structures (Fig. 1b). The cross-like geometry of the cavities in case (ii) is chosen due to its proven efficiency in inducing larger BGs compared to other geometries of the same cavity volume [37]. In both cases, the hierarchical organization is achieved introducing (or repeating) one (or several) self-similar structures obtained by scaling the starting geometry by a hierarchical factor (HF). The representative unit cells of the structures are shown in Fig. 1 with the geometric parameters given in Table 1. In both cases, two HLs of self-similar patterns are considered. The hub-spoke geometry is used to investigate hierarchical elements having similar size scales, i.e. comparable dimensions of the unit cells: $a_1 = 0.4a_2$ (Fig. 1a). The cross-like porous geometry is used to investigate hierarchical elements having an order of magnitude difference in spatial scale: $a_1 = 0.05a_2$ (Fig. 1b). These designs can naturally be extended to $n$ levels of hierarchy.

Bi-periodic metamaterial structures are composed by infinitely duplicating the unit cells in Fig. 1 in a square array. The out-of-plane dimension of the unit cells is assumed to be much larger compared to the lateral ones, allowing to adopt a 2D plane strain approximation [38] [39]. A typical polymeric material such as those used in 3-D printing is considered for the structures. Specifically, we use the mechanical properties of the thermoplastic polymer ABS (Acrylonitrile butadiene styrene), assuming isotropic linear elasticity: bulk modulus $K = 3.34$ GPa, shear modulus $G = 0.714$ GPa, and mass density $\rho = 1050$ kg/m$^3$ [40]. Viscoelastic dissipation is also taken into account, by assuming attenuation linearly proportional to frequency, typical for many polymeric materials [41]. For monochromatic waves, complex-valued elastic constants can then be written, using a Kelvin-Voigt viscoelastic model [42]:

$$K_{ve} = K + i\omega\theta_{loss}, \quad G_{ve} = G + i\omega\eta_{loss}, \tag{1}$$

where the subscript 'vе' indicates the viscoelastic case, $\omega$ is the angular frequency, and $\theta_{loss}$, $\eta_{loss}$ are the viscosity levels. The viscoelastic effects on wave propagation in hierarchically organized metamaterials are discussed in the next section.

Contrary to previous approaches in the literature [32, 27], here the introduction of additional HLs implies an equivalent density reduction, since the aim is to realize lightweight structures with optimized attenuating properties compared to their heavier non-hierarchical counterparts. For the hub-spoke structure, the density reduction is less than 5% compared to the non-hierarchical structure, while for the cross-like porous metamaterial, it reaches 45%.

We consider small-amplitude vector waves in a linear elastic medium. In the absence of external forces the wave field is described by standard wave equations for two-dimensional displacements $\boldsymbol{u}(\boldsymbol{x},t) = \left(u_x(x,y,t), u_y(x,y,t)\right)$ [41]:

$$G\nabla^2 \boldsymbol{u} + \left(K + \frac{G}{3}\right)\nabla(\nabla \cdot \boldsymbol{u}) = \rho \frac{\partial^2 \boldsymbol{u}}{\partial t^2} \tag{2}$$

where $\nabla$ is the gradient operator. The analysis is restricted to harmonic waves and forms a necessary basis for studying other types of wave motions. Due to the structural periodicity of the metamaterials, the displacements are also periodic in space and can be represented using the Floquet-Bloch expansion theorem [16] as follows:

$$\mathbf{u}(\mathbf{x},t) = \mathbf{U}(\mathbf{x})e^{i\mathbf{k}^T \cdot \mathbf{x}}e^{i\omega t} \tag{3}$$

where $\boldsymbol{U}(\boldsymbol{x}) = \left(U_x(x,y), U_y(x,y)\right)$ is the Bloch wave amplitude of the same period as the structural lattice, $\boldsymbol{k} = (k_x, k_y)$ is the wave vector, which is periodic in the corresponding reciprocal space. The Floquet-Bloch theorem (Eq. (3)) allows to restrict the analysis of an infinite periodic medium to that of a representative unit cell only (Fig. 1) with Bloch periodic conditions at the unit cell boundaries:

$$\boldsymbol{u}(x + na, y + ma) = \boldsymbol{U}(x,y)\exp(k_x na + k_y ma) \qquad (4)$$

Here $n, m$ are integers equal to 0 or 1 for an appropriate boundary edge. Non-redundant values of $k_x, k_y$ are located within the unit cell of the reciprocal space – Brillouin zone (Fig. SM1).

The boundary-value problem in Eqs. (2)-(4) can be re-formulated by substituting relation Eq. (3) into Eq. (2), resulting in

$$G\nabla_k^2 \boldsymbol{U} + \left(K + \frac{G}{3}\right)\nabla_k(\nabla_k \cdot \boldsymbol{U}) = -\rho\omega^2 \boldsymbol{U} \qquad (5)$$

where the exponential term $exp(i\omega t+i\boldsymbol{k}^T\cdot\mathbf{x})$ has been divided out $\nabla_k \boldsymbol{U} = \nabla U + i\boldsymbol{k} \otimes \boldsymbol{U}$, and $\nabla_k^2 = \nabla_k \cdot \nabla_k$. The boundary conditions (4) are then replaced by continuity periodic conditions for displacements on the unit cell.

Equation (5) can be solved numerically using the Finite-Element Method (FEM), which allows the analysis of an arbitrarily complex unit cell configuration. The standard Galerkin discretization procedure yields a wave dispersion equation:

$$(\boldsymbol{K} - i(k_x \boldsymbol{K_1} + k_y \boldsymbol{K_2}) + k_x k_y \boldsymbol{K_3} + k_x^2 \boldsymbol{K_4} + k_y^2 \boldsymbol{K_5} - \omega^2 \boldsymbol{M})\boldsymbol{d} = \boldsymbol{0} \qquad (6)$$

or, in shorter form:

$$(\boldsymbol{K_k}(\boldsymbol{k}) - \omega^2 \boldsymbol{M})\boldsymbol{d} = \boldsymbol{0}, \qquad (7)$$

where $\boldsymbol{K}$ and $\boldsymbol{M}$ are standard stiffness and mass matrices, and $\boldsymbol{d}$ is the vector of nodal displacements (see, e.g. [42, 43] for further details) The definitions of $\boldsymbol{K_i}$ matrices are provided in [42]. The values of $\boldsymbol{d}$ are periodic on the unit cell boundaries. The solution procedure to Eq. (7) has been implemented in the Matlab-based code and verified by directly solving Eqs. (2), (4) by means of a modal analysis procedure with the commercial software COMSOL Multiphysics 4.3 [44]. The representative unit cells are discretized into finite elements by using 8-node quadrilateral and triangular elements, respectively. The evaluation of eigenfrequencies $\omega$ has been performed for values of the wave vector $\boldsymbol{k}$ along the borders of the irreducible Brillouin zone, allowing the construction of the corresponding dispersion diagrams (details in the Supplementary Material).

For a viscoelastic medium, the displacement components cannot be represented as a combination of time- and space-dependent functions by using the Floquet-Bloch theorem in Eq. (3). Due to the energy dissipation, spatial wave profiles are time (frequency)-dependent. Also, material characteristics of the medium, $K_{ve}$ and $G_{ve}$, are time-dependent, i.e. $K_{ve} = K_{ve}(t)$ and $G_{ve} = G_{ve}(t)$ and are known as volumetric and shear relaxation moduli, respectively. In the frequency domain, the Floquet-Bloch theorem for a periodic viscoelastic medium takes the form:

$$\widehat{\boldsymbol{u}}(\boldsymbol{x}, \boldsymbol{k}, i\omega) = \widehat{\boldsymbol{U}}(\boldsymbol{x}, \boldsymbol{k}, i\omega) e^{i\boldsymbol{k}^T \cdot \boldsymbol{x}} \tag{8}$$

Here, a hat symbol over the variables indicates the transformed displacements in the frequency domain. The reformulated wave Eq. (5) can then be written as follows:

$$\hat{G}(\omega) \nabla_k^2 \widehat{\boldsymbol{U}} + \left( \widehat{K}(\omega) + \frac{\hat{G}(\omega)}{3} \right) \nabla_k \left( \nabla_k \cdot \widehat{\boldsymbol{U}} \right) = -\rho \omega^2 \widehat{\boldsymbol{U}} \tag{9}$$

where the dependence of vector $\widehat{\boldsymbol{U}}$ on coordinate vector $\boldsymbol{x}$, wave vector $\boldsymbol{k}$, and frequency $\omega$ is implied and omitted for clarity. Similarly to the previously considered elastic case, Eq. (9) can be solved numerically for a representative unit cell subject to continuity periodic conditions for displacements $\widehat{\boldsymbol{U}}$ at the unit cell boundaries.

By using the Galerkin approach, the discretized version of Eq. (9) can be derived:

$$(\widehat{\boldsymbol{K}}_k(\boldsymbol{k}, \omega) - \omega^2 \widehat{\boldsymbol{M}}) \widehat{\mathbf{d}} = \mathbf{0} \tag{10}$$

Here, similarly to Eq. (7), $\widehat{\boldsymbol{K}}_k$ and $\widehat{\boldsymbol{M}}$ are the transformed $\boldsymbol{k}$-dependent stiffness and standard mass matrices of the same form as in Eq. (7), $\widehat{\boldsymbol{d}}$ is the transformed vector of nodal displacements. However, contrary to Eq.(7) and Eq. (10), here the stiffness matrix $\widehat{\boldsymbol{K}}_k$ is frequency-dependent. This dependence does not allow to evaluate eigenfrequencies directly by solving Eq. (10). However, this issue can be overcome by inverting the problem, e.g. by specifying the value of $\omega$ and formulating an eigenvalue problem for wave vector $\boldsymbol{k}$. Since the wave vector has two unknown components, it is necessary to introduce a relation between $k_x$ and $k_y$ in order to derive a typical eigenvalue problem.

Thus, by assuming *e.g.* $k_y = ck_x$ with $c$ a real constant and $0 < c < 1$, Eq. (10) can be rewritten as [43]:

$$\left(\breve{\boldsymbol{K}}_k(\omega) - k_x \breve{\boldsymbol{M}}_k\right)\breve{\boldsymbol{d}}_k = 0 \tag{11}$$

where

$$\breve{\boldsymbol{K}}_k = \begin{bmatrix} -i(\boldsymbol{K}_1 + c\boldsymbol{K}_2) & \boldsymbol{K} - \omega^2 \boldsymbol{M} \\ \boldsymbol{I} & \boldsymbol{0} \end{bmatrix}, \quad \breve{\boldsymbol{M}}_k = \begin{bmatrix} -(\boldsymbol{K}_4 + c\boldsymbol{K}_3 + c^2\boldsymbol{K}_5) & \boldsymbol{0} \\ \boldsymbol{0} & \boldsymbol{I} \end{bmatrix}, \quad \breve{\boldsymbol{d}}_k = \begin{bmatrix} k_x \boldsymbol{d} \\ \boldsymbol{d} \end{bmatrix}.$$

For any specified value of $\omega$, Eq. (11) is an eigenvalue problem for $k_x$ with the continuity periodic conditions for $\breve{\boldsymbol{d}}_k$ at the unit cell boundary. Since most of the available FEM software does not allow to easily modify the element stiffness and mass matrices, the numerical solution procedure for Eq. (7) has been implemented as a Matlab-based code. For a periodic viscoelastic medium, all the eigenvalues $k_x$ are complex-valued with the imaginary part describing spatially attenuated waves, [42].

The evaluation of the wave dispersion for a viscoelastic metamaterial is performed by first fixing the value of $c$ in $k_y = ck_x$ equal to 0 or 1, which corresponds to the $\varGamma X$ or $\varGamma M$ borders of the irreducible Brillouin zone, respectively, and then solving the eigenvalue problem in Eq.(11) for the specific frequencies in the relevant frequency range. Finally, the real parts of the derived solutions $k_x$ are plotted versus the frequencies as common two-dimensional dispersion diagrams. The imaginary parts of $k_x$ are used to evaluate the wave attenuation level $\xi = 2Im(k_x)/Re(k_x)$, thus providing the attenuation spectrum.

To validate attenuation capabilities of the metamaterial structures, we additionally performed transmission analysis on finite-size metastructures with COMSOL Multiphysics 4.3 [45]. The calculations are carried out in the frequency domain for structures composed of 10 adjacent unit cells arranged as a 1D array in the horizontal direction. The structures are modelled as infinite in the vertical direction by applying continuity periodic boundary conditions for wave displacements at the top and bottom boundaries. The analysis is performed using horizontally-polarized plane waves incident on one boundary of the structure to model waves propagating along the $\Gamma X$ direction. The opposite boundary is subject to non-reflecting boundary conditions (Perfectly Matched Layers - PML) to eliminate reflections of the incident wave field. The frequency is swept in 500 or 1000 steps within the ranges specified for each case under consideration from the corresponding dispersion diagrams. The transmitted wave field $T$ is evaluated as:

$$T = log \int_A \left|\frac{\boldsymbol{u}_{tr}}{\boldsymbol{u}_{in}}\right| dA \qquad (12)$$

where $\boldsymbol{u}_{in}$ is the incident displacement amplitude, $\boldsymbol{u}_{tr}$ is the transmitted amplitude, and $A$ is the analysed area, chosen to be representative for the considered wavelengths.

## 3. Linear elastic hierarchical metamaterial

### 3.1. Hierarchical organization at similar spatial scales.

Figure 2 shows dispersion diagrams for the hub-spoke metamaterial (Fig. 1a) with both regular (non-hierarchical) and hierarchical organization. The size of the two representative unit cells $a_2$ is maintained constant, in order to analyse the effects of hierarchy on the wave dispersion in the same frequency range. Direct comparison of the diagrams (Figs. 2a and 2b) reveals two important advantages of hierarchical organization: (i) partial conservation of the BGs originating from the regular geometry and (ii) nucleation of new BGs in the mid-frequency range.

The BGs are said to be partially preserved because the introduction of hierarchy has a twofold effect: it alters some of the pre-existing dispersion curves and it induces new curves, resulting in a possible modification of the original BGs. Physically, this happens because the introduction of the hierarchical architecture alters the stiffness/mass ratio and introduces additional degrees of freedom (at each iteration, a new matrix, a new resonator and new connectors linking the two are added to the system), i.e. new vibration modes are made possible (e.g. see Fig. 2c-f).

The additional bands introduced due to the hierarchical organization can exhibit non-localized or localized behaviour. In the first case, the flat portions of the curves corresponding to localized motions (either at the internal HL1 resonator level (Fig. 2c) or at the ring-like resonator level (Figs. 2d-e)). In the second case, portions of the curves appear with positive or negative slope, involving the combined vibration of both structures occurring in the hierarchical geometry (Fig. 2f). Since these curves can be located inside or outside a pre-existing BG at the previous HL, they can cause the BG split into two or more BGs (compare grey rectangles in Figs. 2a and 2b).

Note that the upper and lower edges of the original BGs are left unaltered by the introduction of the hierarchical structure. The analysis of the edge mode shapes of the BGs in both the original and the hierarchical structures (Figs. 2g-o) unveils the physical reason for this: in both cases, the lower and upper bounds of the original BGs involve strong motion that is highly localized in the HL2 matrix, and therefore the introduction of the nested structure HL1 does not alter the vibration shapes and frequency values of the BG bounds (Figs. 2g-h, j-k, l-m and n-o).

Figures 3a-c show that the frequencies of hierarchically-induced bands can be directly tuned by varying the HF value due to an increase or decrease of the mass/stiffness ratio of the structural elements [46]. For instance, increasing the HF from 0.38 to 0.42 shifts the HL1 rotational mode (see Fig. 2c for its mode shape) highlighted in green in Figs. 3a-c towards lower frequency values from 639 kHz to 581 kHz. This occurs since by increasing HF, so the HL1 resonator becomes larger and the added mass results in lower vibration frequency. A similar behaviour is observed for the other two considered bands.

Another feature introduced by hierarchical organization is the nucleation of new BGs. The diagram for the nested hierarchical unit cell has two additional BGs at around 1 and 1.4 MHz, respectively, highlighted with red rectangles in Fig. 2b. To explain the physics of this novel BG mechanism, we observe that they do not originate from a simple superposition of the dispersion spectra for the two constitutive structures. This is evident by considering the diagram in Fig. 2a and its scaled equivalent obtained by multiplying frequencies by a factor of $1/HF = 2.5$, so that the first BG between 450 and 803 kHz is shifted to frequencies between 1.125 - 2.0 MHz. Hence, simple unit cell scaling is not responsible for the presence of the two new BGs. Further, the bounds of the additional BGs are *not* at the high-symmetry points of the first irreducible Brillouin zone's border and are *not* described by completely flat curves representing localized modes (although some flat regions are present). Hence, the BG nucleation can neither be attributed to pure Bragg scattering nor to pure local resonance mechanisms. Calculation of the vibration modes at the edges of the new BGs (Fig. 3d-f) reveals a novel deformation mechanism in which all the HLs are involved. Specifically, the upper bounds of the first new BG (at approximately 1 MHz) involves the deformation of the "ring" element introduced by hierarchy (see Figs. 3d-f, for $HF = 0.38, 0.40$ and $0.42$, respectively) and of the ligaments which undergo bending or extensional deformations, in some cases with opposite phases . The greater the HF is, the more the motion is localized at the HL1, limiting the deformations in the HL2 matrix (see Figs. 3d-f). A similar mechanism (with different motion distribution and deformation) takes place at the second new BG. The HF significantly influences the BG width, too: decreasing the HF from 0.42 to 0.38, it is possible to obtain up to approximately 3 times larger BGs (see Fig. 3a-c). Given the critical role played by hierarchy in the nucleation of these new BGs, they are referred to as "hierarchically induced BGs".

The comparison of the band structures in Figs. 2*a,b* highlights another advantage of hierarchical organization: the similarity of dispersion curves below the first BG frequencies with those of the regular structure. This occurs since low-frequency waves (below 250 kHz for the analysed structure) are insensitive to the absence of small portions of the material due to the introduction of the hierarchy. However, at higher frequencies, from 250 kHz to 450 kHz, the similarity of the dispersion curves is preserved only in the case of hierarchical structures displaying self-similarity (see Fig. SM5 in the Supplementary Material). Starting from the BG frequencies, the number of modes for the hierarchical metamaterial is higher compared to that for the regular structure. This is mainly due to the presence of additional localized modes inside the internal hierarchical object (flat curves in the diagram). For example, a flat curve at 622 kHz corresponds to a torsional mode of the inclusion of HL 1 (Fig. 2*e*). Finally, hierarchical structure allows to isolate dispersion curves with negative group velocity (see Figs. 3a-c), which also makes hierarchically organized metamaterials attractive for potential applications exploiting negative refraction [47, 48].

### *3.2. Hierarchical organization at different spatial scales.*

Next, we analyse wave dispersion in a cross-like porous metamaterial with an external hierarchical organization at different spatial scales (Fig. 1b). The band structures for the regular and hierarchical metamaterial configurations are shown in Figs. 4*a,b*, respectively. It is apparent that the wide BG ranging from approximately 300 kHz to 740 kHz in Fig. 4*a* is almost completely preserved in the diagram of Fig. 4*b*. Pass bands located inside the latter BG are represented by mostly flat lines corresponding to localized modes (Fig. 4*c*). As discussed below, these modes become evanescent in realistic structures, which always have a certain level of energy dissipation.

Figure 5 shows the dispersion diagrams for the unit cell of a regular metamaterial of the size of HL 2 (a) and hierarchical metamaterial (b) in a low-frequency range. It can be seen that by introducing the hierarchy the lowest BG is shifted to approximately 3 times lower frequencies, while several other BGs appear at the BG frequencies of the regular structure. The first observation to be made is that the sub-wavelength BG is opened in a more lightweight hierarchical structure, a feature of hierarchically organized metamaterials that can provide the possibility for the design of lightweight metamaterials to manipulate low-frequency waves. The other BGs are of Bragg type, induced by the changes in the unit cell geometry due to hierarchical structure, leading to increased multiple scattering of elastic waves. Also, due to the self-similarity of the constitutive geometries at different HLs, the dispersion curves below the lowest BG have similar trends for hierarchical and non-hierarchical structures, together with the corresponding vibration patterns (Fig. 5 *c-d*).

To explain the mechanism of the sub-wavelength BG generation, we consider an equivalent mass-spring model capable of predicting the bounds of the first BG in a metamaterial with cross-like holes [37]. According to this model, a square lattice of cross-like cavities can be considered as a lattice of square blocks connected by thin ligaments (Fig. 6a). At low frequencies, the blocks vibrate as rigid bodies; whereas ligaments play the role of springs (Fig. 6b). Thus, the lower bound of the BG can be evaluated as [37]:

$$\omega_l^{BG} = \sqrt{\frac{\overline{K}}{\overline{M}}}, \qquad (7)$$

where $\overline{K} = 2S\left(\frac{\lambda + 2\mu}{l_b + l_s} + \frac{\alpha\mu}{l_s}\right)$ and $\overline{M} = m_l + \frac{m_s}{2}$ are effective stiffness and effective mass, respectively; $m_b, m_s$ and $l_b, l_s$ are masses and lengths of the block (subscript $b$) and springs (subscript $s$); $S$ is the cross-section of the springs, $\lambda = K - \frac{2}{3}G$ and $\mu = G$ are the Lamé constants, and $\alpha$ is a weighting coefficient.

The same model can be applied to the considered self-similar hierarchical structure. Obviously, by introducing the hierarchy, the lengths of the blocks and ligaments remain unchanged. Therefore, by evaluating the effective stiffness $\overline{K}^h$, the only variable quantity is the cross-section $S_1^h = (a_1 - 2b_1) \times 1$, which is $S_1^h = HF \cdot S_2^h$, with $S_2^h$ indicating the cross-section of the structure at HL 2 (superscript $h$ refers to a hierarchical structure). As mentioned previously, the mass of the hierarchical unit cell is 45% of the regular counterpart, i.e. $\overline{M}^h = 0.45\overline{M}^r$ (superscript $r$ refers to a regular structure). The ratio of the lower bounds for the band gaps in the hierarchical and regular structures is then as follows:

$$\frac{\omega^h}{\omega^r} = \sqrt{\frac{\overline{K}^h \cdot \overline{M}^r}{\overline{M}^h \cdot \overline{K}^r}} = \sqrt{\frac{0.05}{0.45}} = \frac{1}{3} \qquad (8)$$

Thus, the equivalent spring-mass model adapted for a hierarchical structure with cross-like holes predicts a shift of the lower BG bound to 3 times lower frequencies, while the calculated data gives $\frac{\omega^h}{\omega^r} = 2.82$, i.e. the accuracy of the model approximation is 94%. One may conclude that the lowest BG in the band structure of the hierarchical unit cell is of the same type as that for a regular unit cell, and its shift to sub-wavelength frequencies occurs due to hierarchical structure.

Similar shifts can be achieved by exploiting hierarchical organization of non self-similar structures. However, self-similarity of hierarchical constituents enables to preserve similarity in the dispersion curve trends below the lowest BG. Above the BG, the curve trends differ, but the vibration patterns in the vicinity of the BG are still similar (see Fig. 5 *e-f*).

Overall, the external hierarchical organization can be considered as a means of tuning BGs to lower frequencies accompanied by the simultaneous reduction of the structural weight and generation of additional BGs at higher frequencies. This mechanism may extend applicability ranges of mechanical metamaterials and facilitate their wide practical utilization. Moreover, the three main advantages found for the nested hierarchical organization at similar spatial scales are preserved for the metamaterial with the external hierarchy at different spatial scales. Specifically, these are (1) (partial) conservation of the BGs induced by the constitutive geometries, (2) nucleation of additional BGs in the mid-frequency range, and (3) similarity of the dispersion curves below the BG frequencies for the self-similar hierarchical structure and the regular counterpart. These properties are common to various hierarchical structures, as demonstrated in the Supplementary Material, where other designs are considered with similar results.

## 4. Dissipative hierarchical metamaterials

Many materials, e.g. polymers used to produce 3D-printed samples, typically display wave dissipation characteristics. For solid structures, this can be taken into account by considering viscoelastic properties of the metamaterial constituents [49, 50]. Viscoelastic material behaviour usually influences wave propagation characteristics of mechanical metamaterials and modifies band structure diagrams calculated under assumption of the linear elasticity [42].

To analyse viscoelastic effects on wave dispersion in hierarchically organized metamaterials, we assume dissipation to be linearly proportional to frequency, which is typical of many polymers [41, 49]. Since, under isothermal conditions, the bulk modulus of most polymers appears to retain a stationary equilibrium value over a range of time scales [42, 49, 51], i.e. $K_{ve} = K$ and $\theta_{loss} = 0$ in Eqs. (1), time-dependency of the shear modulus $G$ can be effectively exploited. In the frequency domain, viscoelastic dissipation is then described by the Kelvin-Voigt model presented by Eqs. (1), which well approximates the response of real polymers [42] and are commonly used to study damped metamaterials [42, 43].

To model physically realistic scenarios, the values of shear viscosity $\eta_{loss}$ are chosen by considering the viscoelastic effect on the wave dispersion in a homogeneous material, similarly to [42]. It has been found that for $\eta_{loss} \leq 5$ Pa·s, the real part of wave vectors for propagating modes is almost unchanged from that for the corresponding elastic case, while the imaginary part shows small deviations from zero values at higher frequencies in the considered frequency range, up to 800 kHz. Hence, these values correspond to small material dissipation [42, 52]. Considering that the objective of this study is not to model wave dissipation in any specific polymer, but rather to theoretically analyse the influence of viscoelasticity on the dynamics of hierarchically organized metamaterials, we choose the representative values of $\eta_{loss} = 2$ Pa·s and $\eta_{loss} = 5$ Pa·s to obtain indicative results.

For a viscoelastic medium, pure propagating waves or BGs can no longer be identified, since all waves are attenuating or evanescent. Thus, the study is performed for experimentally realistic scenarios, for which waves are characterised by real-valued frequencies and complex-valued wave vectors $\mathbf{k} = \text{Re}\,\mathbf{k} + i\,\text{Im}\,\mathbf{k}$ with the imaginary part $\text{Im}\,\mathbf{k}$ describing the wave spatial attenuation. The intensity of the wave attenuation can be evaluated by considering either an attenuation spectrum relating the frequency to the wave attenuation level $2\mathrm{Im}(\mathbf{k})/\mathrm{Re}(\mathbf{k})$ [42, 50] or transmission spectrum showing non-normalized amplitudes of transmitted waves versus the wave frequency.

Figure 7a shows a dispersion diagram for waves propagating in the $\Gamma-X$ direction in the hub-spoke hierarchical metamaterial together with attenuation (Fig. 7*b*) and transmission (Fig. 7*c*) spectra for elastic and viscoelastic ($\eta_{loss} = 2$ Pa·s) material behaviour. Figure 7*a* reproduces a part of Fig. 2b for only one propagation direction, with the corresponding number and sizes of the BGs. The BGs for the elastic structure are highlighted by shaded regions. According to the definition, the wave attenuation for propagating modes ($\mathrm{Im}\mathbf{k}=0$) in the elastic case is zero outside the BGs and has finite values for the evanescent modes ($\mathrm{Im}\mathbf{k}\neq 0$). The attenuation diagram of waves shown in Fig. 7*b* is in good agreement with their transmission characteristics for both elastic and viscoelastic structures, discussed below. Despite the small level of viscosity, wave attenuation/transmission in the viscoelastic metamaterial differs from that in its elastic counterpart starting from frequencies of the lowest BG. In general, the attenuation (transmission) increases (decreases) linearly with the frequency, in agreement with the adopted assumption. Viscoelasticity at most influences the BG bounds, where straight edges of attenuation curves, typical for the elastic case, become "rounded" in a similar way as for regular metamaterials [42, 52]. No peculiar viscoelastic effects are found for the additional BGs opened due to hierarchy, presumably due to the fact that their width is very narrow for this metamaterial geometry.

A similar behaviour is found for low-frequency wave attenuation in the cross-like porous metamaterial with external hierarchy at different spatial scales. Figure 8a shows a part of the dispersion diagram given in Fig. 4c for waves propagating in the $\Gamma-X$ direction in the elastic cross-like porous hierarchical metamaterial. In the hierarchical structure, many BGs exist, separated by almost flat horizontal pass bands, at the BG frequencies of the regular metamaterial counterpart (Fig. 4*a*). The multiple pass bands found in this spectrum can also be observed in the transmission spectra represented by the averaged surface displacement (Fig. 8b). Transmitted displacements over 1 unit cell, which is the closest to the loaded boundary (black curve in Fig. 8b), do not reveal the presence of any BG at all, whereas at the distance of 5 unit cells from the loading (blue curve in Fig. 8b), clear transmission dips at the BG frequencies are observed.

Thus, when small dissipation is added, all the pass bands are transformed into attenuating waves with large values of the imaginary part, as can be clearly seen from the attenuation spectra for the elastic and viscoelastic metamaterials in Fig. 8c. Here, attenuation for propagating elastic modes equals zero and is given in blue, while attenuation of evanescent modes for elastic and dissipative material behaviour given in black and green, respectively. Thus, the introduction of viscoelasticity leads to a considerable modification of the attenuation spectrum. The curves do not resemble those for the elastic case any longer, and all modes become strongly attenuating. Maximum attenuation occurs at lower frequencies, since the modes near 555 kHz form the upper bound of a large BG for the elastic metamaterial, where curve "rounding" effect occurs [52]. The same conclusion can be derived from the analysis of the transmission. As can be seen in Fig. 8d, 1 unit cell of dissipative hierarchical metamaterial is sufficient to reduce the amplitude of an incoming wave up to 4 orders of magnitude, and at the distance of 3 unit cells a reduction of up to 10 orders can be achieved. Hence, taking into account viscoelastic material behaviour of hierarchically organized metamaterials allows not only to obtain a more realistic mathematical model and results for these structures, but also shows an advanced wave attenuation performance of the metamaterial.

Similar effects to those described are observed in both considered hierarchical metamaterials when the viscosity level is increased to $\eta_{loss} = 5$ Pa·s. This shows that the influence of viscoelasticity on wave dissipation in hierarchical metamaterials is not necessarily related to the chosen viscosity levels and the structural organization plays an important role.

In summary, realistic modelling of wave dispersion in hierarchical metamaterials, taking into account damping effects present in any real material, shows that hierarchical organization at various spatial scales enables to completely preserve BGs induced by constitutive hierarchical objects.

## 5. Conclusions

We have numerically investigated the influence of bio-inspired hierarchical organization and material viscoelasticity on wave dispersion in metamaterials with self-similar constituents at various spatial scales. Contrary to previous approaches, our study focuses on porous hierarchical structures, whereby increasing hierarchy entails a weight reduction. Results reveal three main advantages of the hierarchical structure on the dynamic performance of mechanical metamaterial. These are:

(1) conservation of most of the BGs induced by the constitutive regular geometries, in the presence of material damping;

(2) nucleation of additional "hierarchically-induced" BGs in the mid-frequency range;

(3) similar wave dynamics at low frequencies for hierarchical and corresponding regular structures.

We have discussed the physical mechanism for the occurrence of hierarchically induced BGs, and shown the generality of the observed behaviour, which is not limited to a specific configuration. Additionally, our results demonstrate that an "external" structural hierarchy occurring at different spatial scales can be exploited to tune BGs of the regular metamaterial counterpart to lower frequencies while significantly reducing structural weight. A simple equivalent mass-spring model has been developed to predict this BG shift. We have analysed for the first time the effect of material dissipation losses on wave propagation in hierarchical metamaterials, establishing their crucial role, even when they are small, on the preservation of the BG size. Overall, this work provides insights on the crucial role of hierarchical structure on the dynamic behaviour of metamaterials and reveals that the principles of bio-inspired hierarchical organization can lead to lightweight metamaterials with advanced multi-frequency attenuation properties, providing useful design principles for practical applications.

**Acknowledgements**

M. M. acknowledges funding from the European Union's Horizon 2020 research and innovation programme under the Marie Skłodowska-Curie grant agreement N. 658483. A. K. acknowledges funding from the European Union's Seventh Framework programme for research and innovation under the Marie Skłodowska-Curie grant agreement N. 609402-2020 researchers: Train to Move (T2M). N.M.P. is supported by the European Research Council PoC 2015 "Silkene" No. 693670, by the European Commission H2020 under the Graphene Flagship Core 1 No. 696656 (WP14 "Polymer Nanocomposites") and FET Proactive "Neurofibres" grant No. 732344. F.B. is supported by H2020 FET Proactive "Neurofibres" grant No. 732344.

**Figure Captions**

Figure 1: Hierarchical "porous" structures found in natural materials at different spatial scales (in this example in diatom cell walls [53] [54]) and the corresponding bioinspired hierarchical metamaterial unit cells: (a) structures with thin connecting elements and nested hierarchical organization (hub-spoke geometry) and (b) cross-like cavities and external hierarchical organization (cross-like porous metamaterial).

Figure 2: Dispersion diagrams for the hub-spoke metamaterial with: (a) regular and (b) hierarchical organization. The size of the two representative unit cells is maintained constant and equal to $a_2$, in order to analyse the effects of hierarchy on the wave dispersion in the same frequency range. Ordinary BGs are indicated in grey. Hierarchical structure allows to induce additional BGs at mid- and high-frequencies (highlighted in red). (c)-(o) Vibration patterns at various frequencies. Coloured symbols indicate their corresponding location on the band diagrams.

Figure 3: Dispersion diagrams for the hub-spoke metamaterial for three different values of the HF: (a) 0.38, (b) 0.40 and (c) 0.42. (d-f) Mode shapes at the edges of the "hierarchically-induced" BGs.

Figure 4: Dispersion diagrams for the cross-like porous metamaterial with (a) regular and (b) external hierarchical organization. The size for the unit cell of the regular structure is the same as that for hierarchical level 1. (c) A part of the dispersion diagram for a hierarchical metamaterial inside the BG. The hierarchical organization at different spatial scales allows preserving the BG originating from a smaller constitutive geometry.

Figure 5: Dispersion diagrams for the cross-like porous metamaterial with (a) regular and (b) hierarchical organization. The size for the regular unit cell is the same as for HL 2. The hierarchical organization allows shifting of the lowest BG to sub-wavelength frequencies. (c)-(f) Vibration patterns for the regular and hierarchical unit cells.

Figure 6: (a) Four adjacent unit cells of the regular cross-like porous metamaterial and an equivalent spring-mass model capable of evaluated the lower bound of the 1st BG [28]. (b) The vibration pattern of the lower bound of the 1st BG.

Figure 7: (a) Dispersion diagram for waves propagating in Γ-X direction in the hub-spoke metamaterial with hierarchical organization. (b) Attenuation spectrum for the same metamaterial with linear elastic $(K, G)$ and viscoelastic $(K_{ve} = K, G_{ve} = G + 2i\omega)$ material behaviour. (c) Transmission spectrum for elastic and viscoelastic hierarchical metamaterials. Propagating elastic waves are shown by blue, attenuating elastic and viscoelastic waves are shown by black and green, respectively. Viscoelasticity leads to the expected frequency-proportional increase of the wave attenuation with a "rounding" of the curve trends near the BG bounds.

Figure 8: (a) Dispersion diagram for waves propagating in Γ-X direction in the cross-like porous metamaterial with hierarchical organization. (b) Transmission spectra for the 1st (black) and last 5 (blue) unit cells of the elastic finite-size metamaterial structure. (c) Attenuation spectrum for the same metamaterial with elastic $(K, G;$ black$)$ and viscoelastic $(K_{ve} = K, G_{ve} = G + 2i\omega;$ green$)$ material behaviour. (d) Transmission spectra for the 1st (light green) and medium 5 (dark green) unit cells of the viscoelastic finite-size metamaterial structure.

**Table captions**

Table 1: Case studies and geometrical parameters used in the simulations. **HF** is a **H**ierarchical **F**actor indicating the ratio between hierarchical levels n and n+1, i.e. **HF** = $a_n/a_{n+1}$.

**Figures**

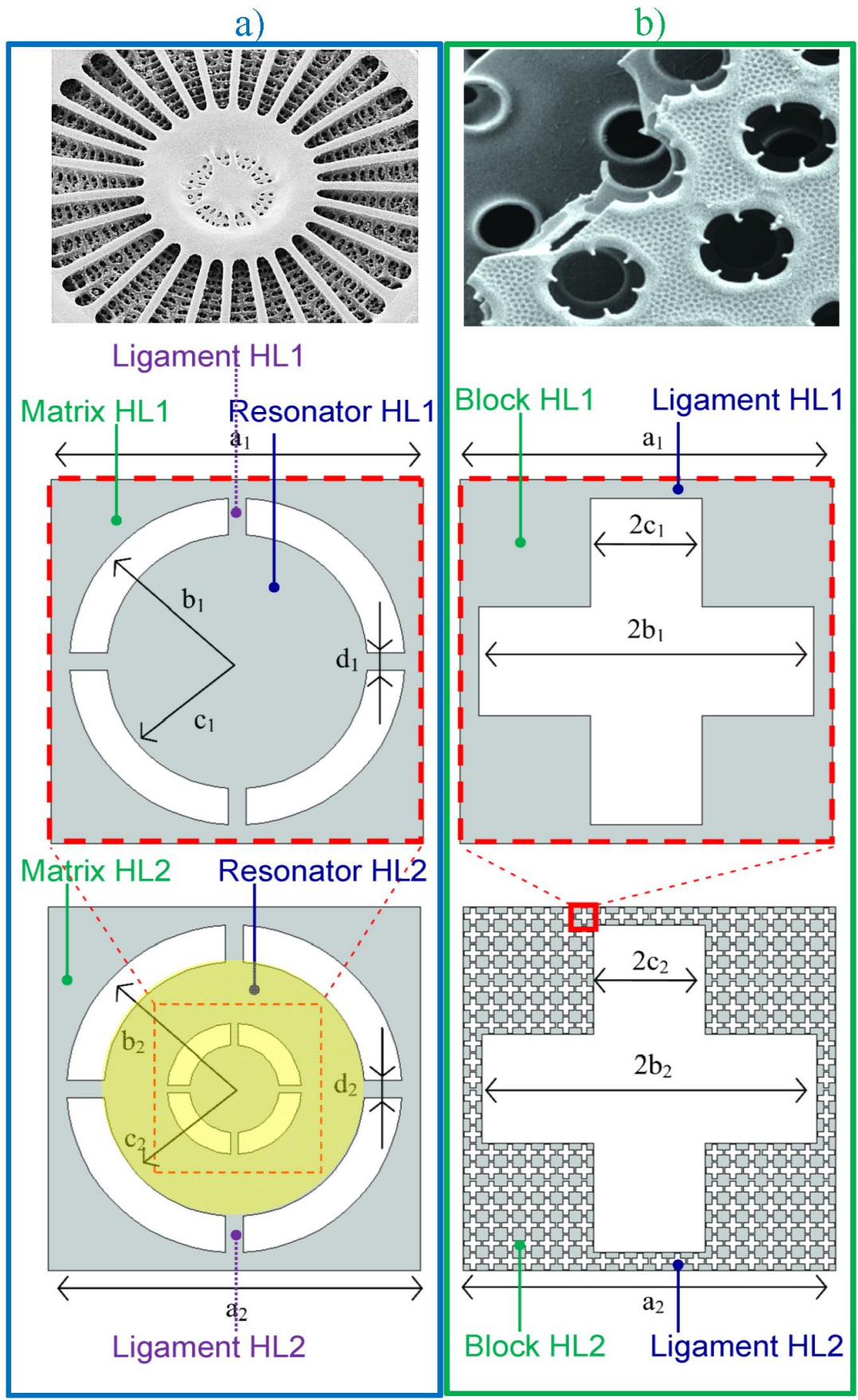

*Figure 1*

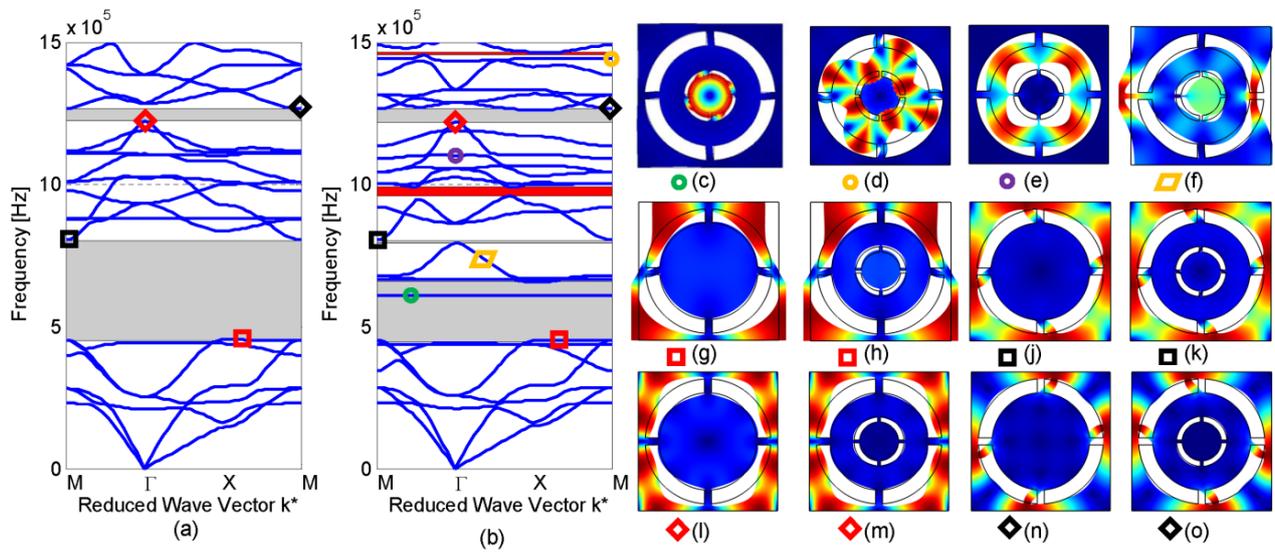

*Figure 2*

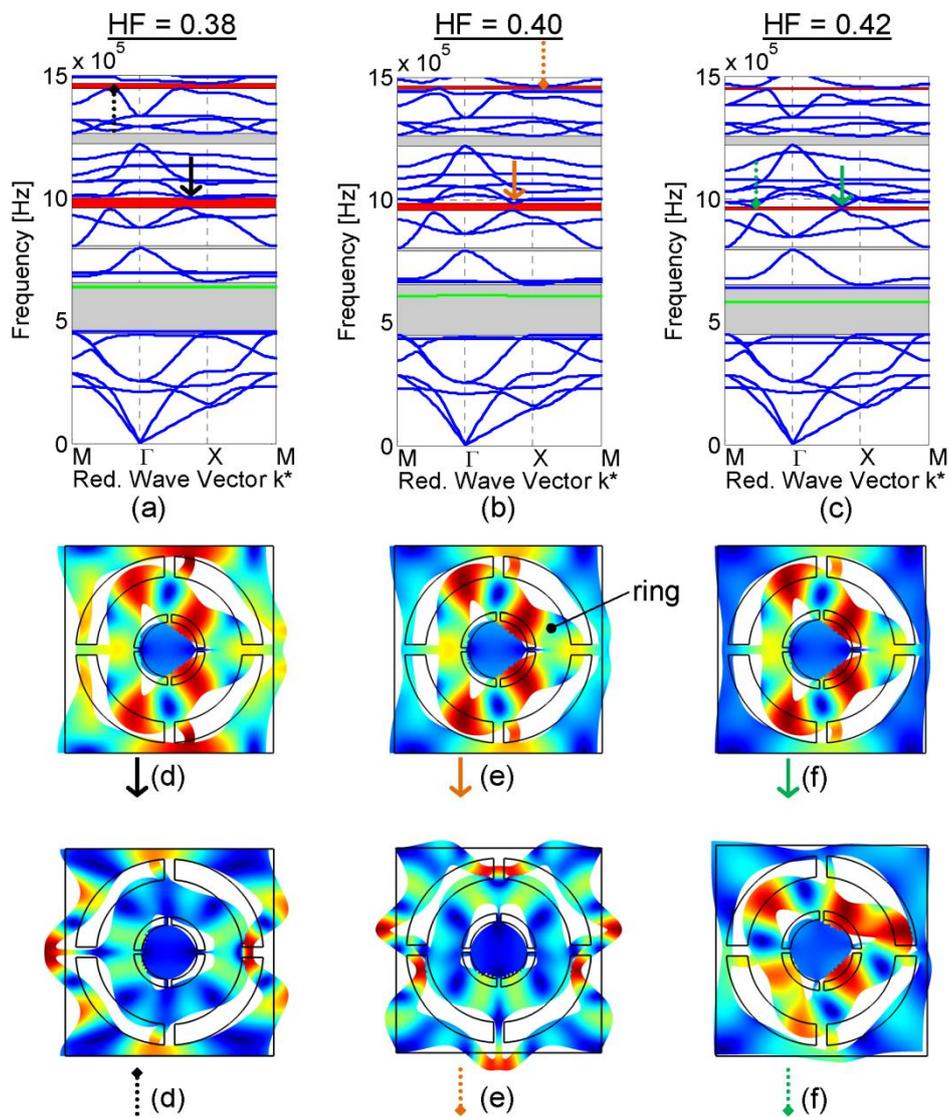

*Figure 3*

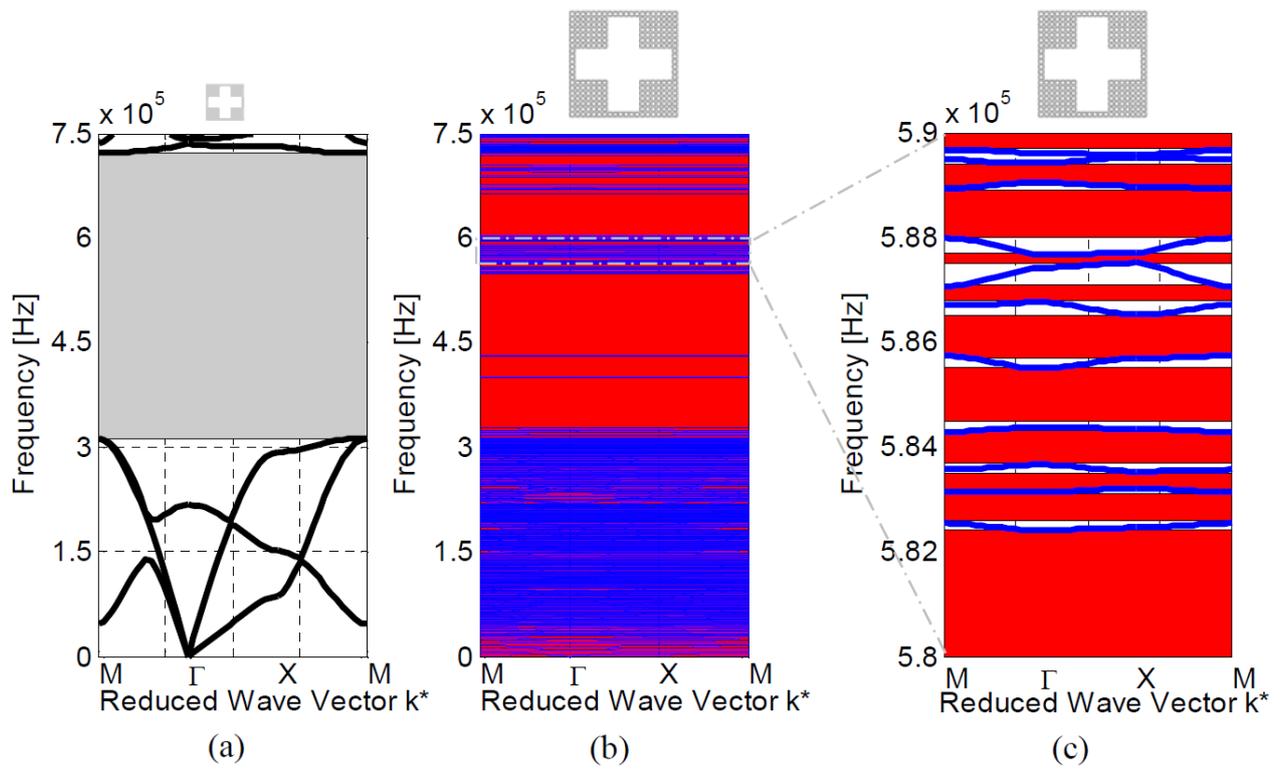

*Figure 4*

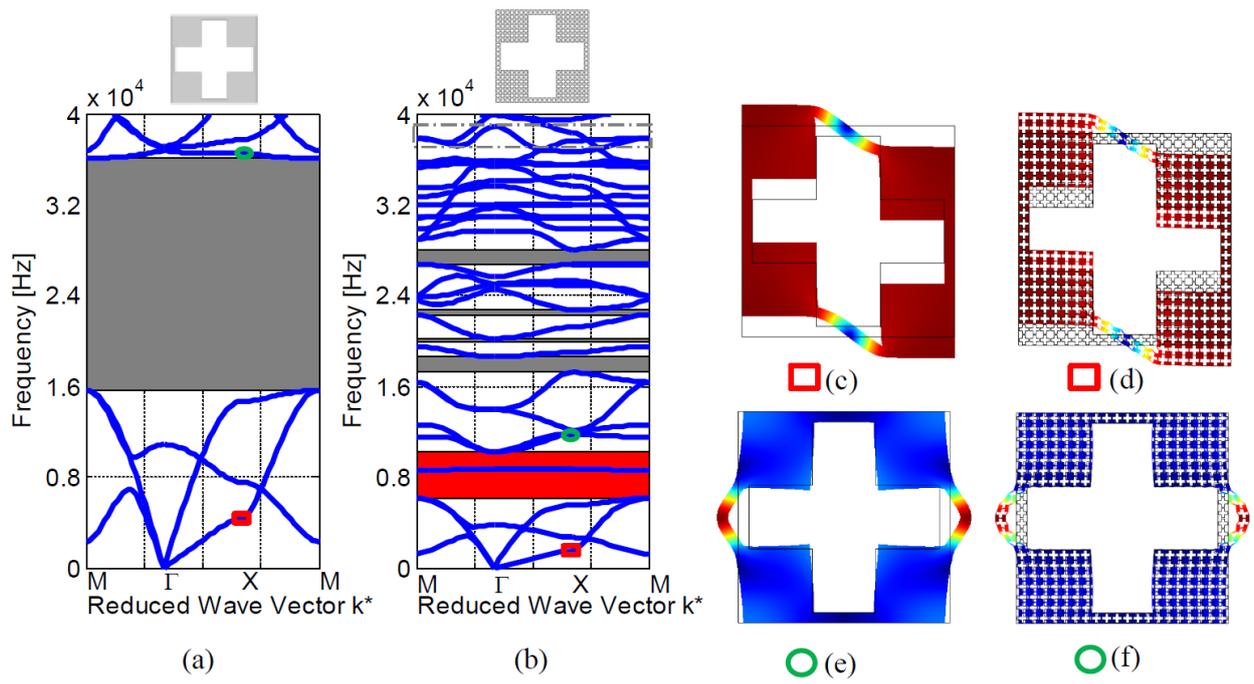

*Figure 5*

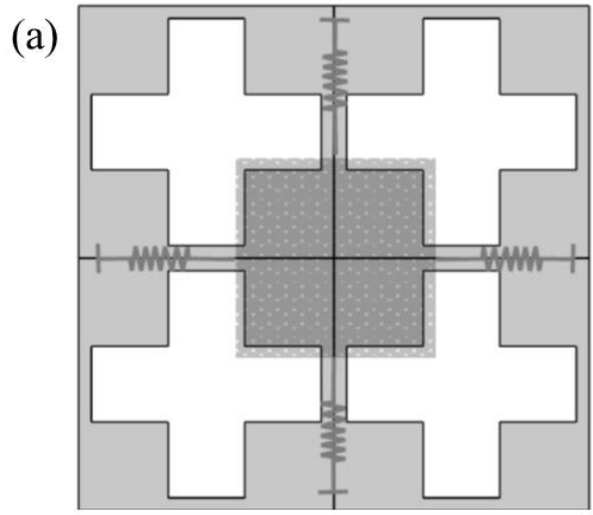

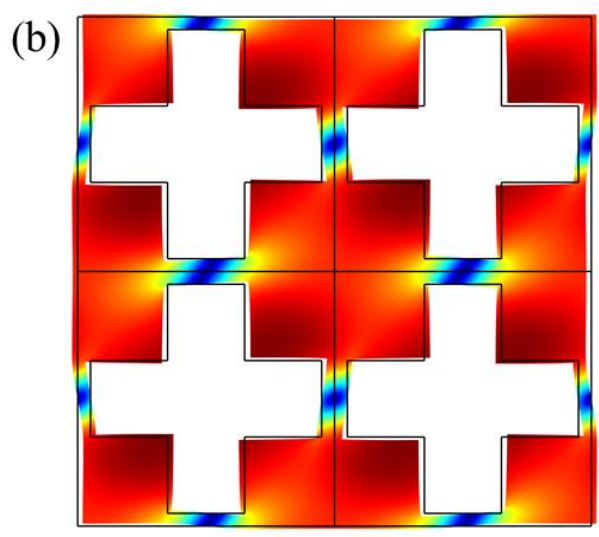

*Figure 6*

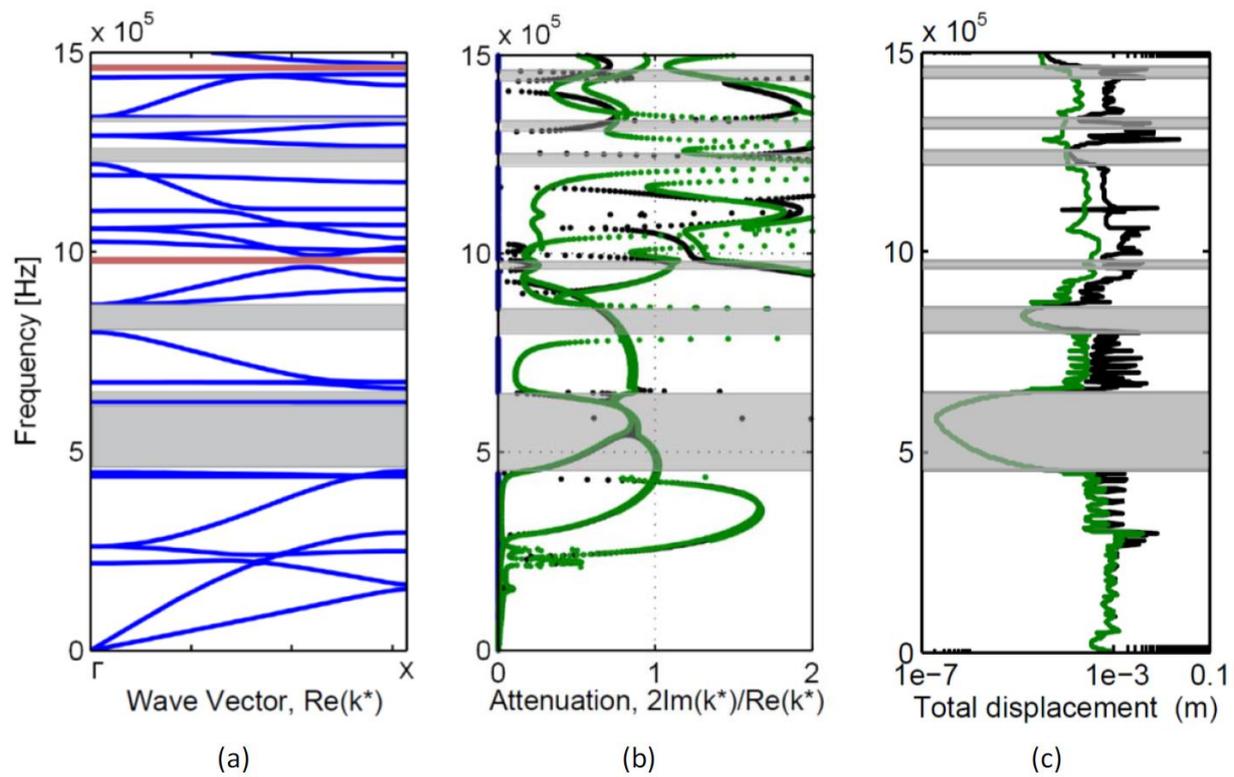

*Figure 7*

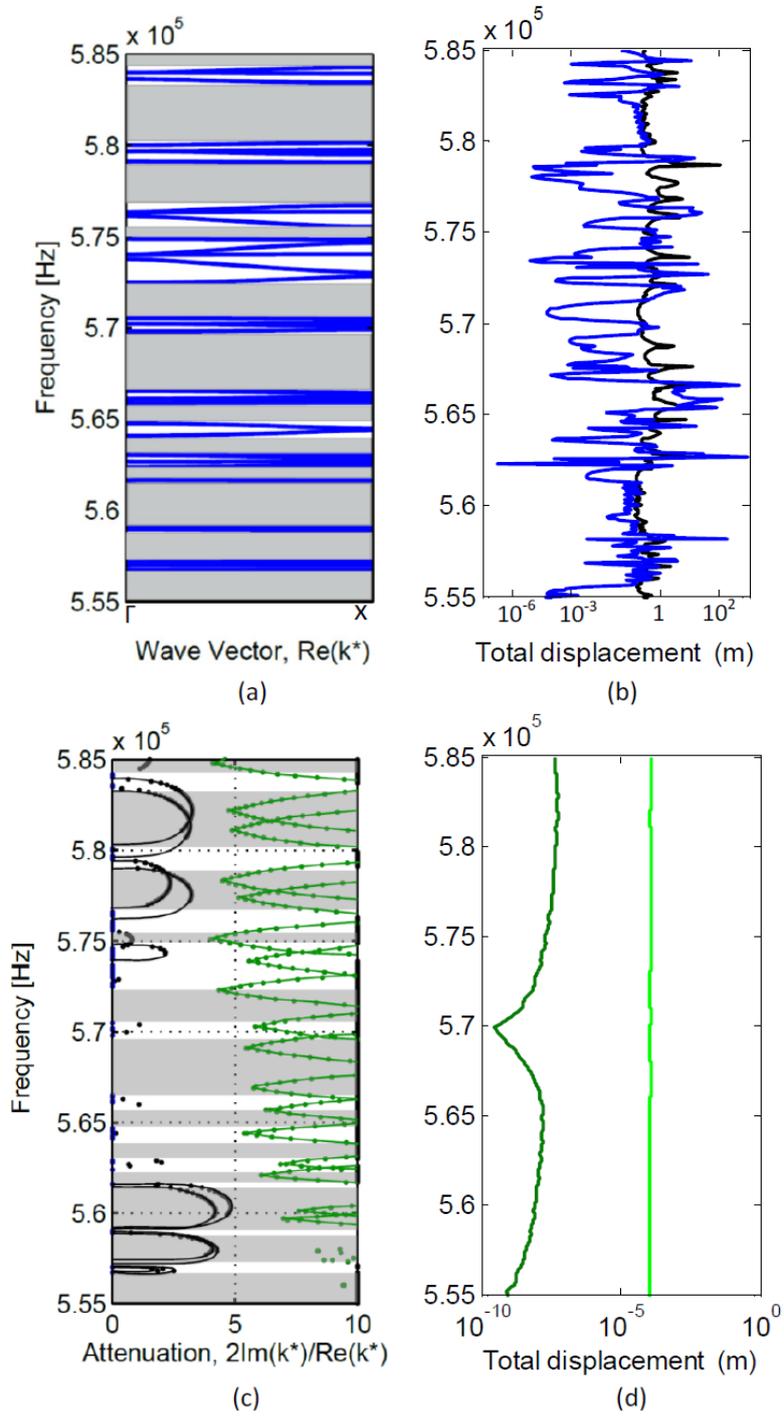

*Figure 8*

**Tables**

| Case study | *HF* | Parameters [m] | | | |
|---|---|---|---|---|---|
| | | $a_n$ | $b_n$ | $c_n$ | $d_n$ |
| Hub-spoke | 0.4 | $4 \cdot 10^{-4}$ | $0.45 a_n$ | $0.35 a_n$ | $0.05 a_n$ |
| Cross-like porous | 0.05 | $1 \cdot 10^{-3}$ | $0.45 a_n$ | $0.15 a_n$ | - |

*Table 2*